\documentclass[dvips]{article}

\usepackage{icrc2011}

\title{Prospects of performing Lorentz invariance tests with VHE emission from pulsars}

\shorttitle{Nepomuk Otte Prospects of performing Lorentz invariance tests with pulsed VHE emission from Pulsars}

\authors{A.\ Nepomuk\ Otte$^{1}$}
\afiliations{$^1$SCIPP, University of California, 1156 High Street, Santa Cruz, CA 95060, U.S.A.}
\email{nepomuk.otte@gmail.com}

\abstract{Gamma-ray observations provide sensitive tests of Lorentz invariance violation (LIV). At present the most sensitive tests come from observations of transient events, gamma-ray bursts and flaring AGN. Disadvantages of transients are that an independent confirmation by a different experiment is often not possible and that limits cannot be improved with a longer exposure. Pulsars do not have these disadvantages. Testing Lorentz invariance with pulsars was not considered seriously so far because limits were not competitive. The VERITAS collaboration has recently reported the detection of pulsed emission from the Crab pulsar above 100\,GeV. This measurement can be used to constrain LIV effects with a sensitivity that is competitive with some of the best available limits. In view of this unexpected result we discuss what the prospects are of doing LIV tests with very-high energy gamma-ray emission from pulsars. }

\begin{document}
\maketitle

\begin{figure*}[htb]
  \centering
  \includegraphics[width=4.5 in]{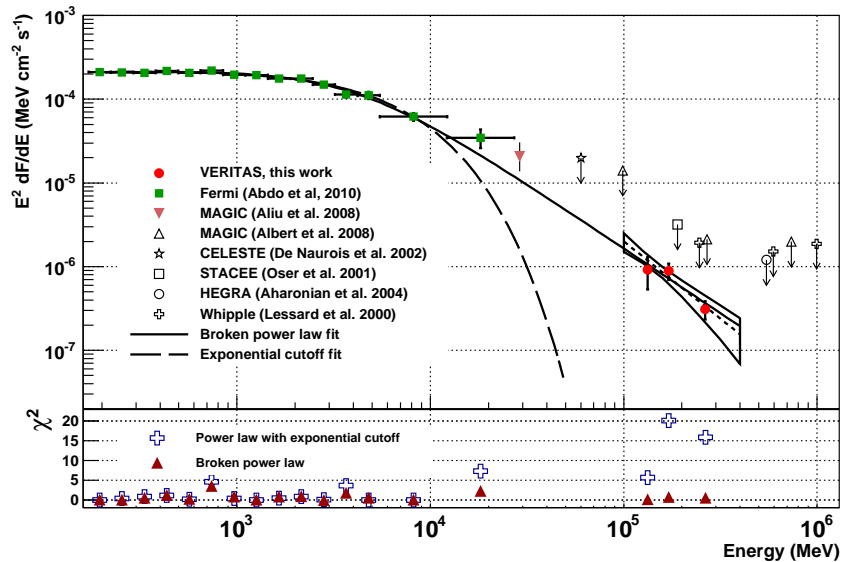}
  \caption{Spectral energy distribution (SED) of the Crab pulsar in gamma rays. Figure from \cite{andrew}. }
  \label{sed}
 \end{figure*}

\section{Introduction}

Lorentz invariance (LI) is a fundamental concept of modern physics. One of the consequences of LI is that the speed of light is constant and, in particular, that it does not depend on the photon energy. If, however, spacetime has structure, like it is postulated in some approaches to combine quantum mechanics and general relativity \cite{Smolin, Amelino:98},
 it could be that the speed of light depends on the energy of the photon. From our everyday experience we know that the effect, if it exists, is very small and, therefore, only evident if the energy of the photons is in the gamma-ray regime and the photons have traveled large distances. In conclusion, Lorentz invariance violation (LIV) can be tested effectively by searching for a delay in the arrival time of gamma rays with different energies that have been emitted simultaneously from an astrophysical object. 

A quantitative prediction of the energy dependency of the speed of light does not exist. A pragmatic ansatz that is often made is to modify the constant speed of light $c$ by adding energy dependent terms that are proportional to $(E/E_{\rm LIV})^n$, where $E$ is the energy of the photon, and $E_{\rm LIV}$ can be considered as the energy scale at which Lorentz invariance violating effects become evident. $E_{\rm LIV}$ is the quantity that is normally tested in LIV tests. With this ansatz, two photons with different energies $E_h$ and $E_l$ that are emitted simultaneously arrive at an observer at a distance $d$ at slightly different times. Depending on the order $n$ of the energy dependence, the time difference $\Delta t$ for the linear and the quadratic term in $E/E_{\rm LIV}$ are:
\begin{equation}\label{lin}
\Delta t_1 = \frac{d}{c} \cdot\frac{E_{\rm h} - E_{\rm l}} {E_{\rm LIV}}\rightarrow E_{\rm LIV} =  \frac{d}{c} \cdot\frac{E_{\rm h} - E_{\rm l}} {\Delta t_1}
\end{equation}
and
\begin{equation}\label{quad}
\Delta t_2 = \frac{d}{c}\cdot \frac{3}{2} \cdot \frac{E^2_{\rm h} - E^2_{\rm l}}  {E^2_{\rm LIV}}\rightarrow E_{\rm LIV} =\sqrt{\frac{d}{c}\cdot \frac{3}{2} \cdot \frac{E^2_{\rm h} - E^2_{\rm l}}  {\Delta t_2}}\,, 
\end{equation}
respectively.

In Section 2 I discuss the status of LIV tests with flaring AGN and GRB, which provide the most stringent tests of $E_{\rm LIV}$ to date. In Section 3 the advantages of doing LIV tests with pulsars are explained. In Section 4 I give a quantitative estimate of how much LIV can be constrained with the recent detection of the Crab pulsar at 120\,GeV by the VERITAS collaboration \cite{andrew}. In Section 5 I discuss the prospects of doing LIV tests with pulsars. The paper closes with summarizing remarks in Section 6. 

\section{LIV tests with GRB and AGN}

At present the most stringent tests of LIV come from the observation of flaring active galactic nuclei (AGN) \cite{HESS} and gamma-ray bursts (GRB) \cite{Fermi} in the gamma-ray band. AGN and GRB are extragalactic gamma-ray sources, which is advantageous because the time delay $\Delta t$ depends linearly on the distance. A large distance is most helpful in constraining the linear case, cf.\ Equation \ref{lin}. In case of constraining higher order terms, cf.\ Equation \ref{quad}, it becomes more important to detect photons at higher energies. This illustrates GRB GRB090510 that went off at a distance of z=0.903 and provides the strongest constraint on the linear term based on the detection of one 31 GeV photon 0.829 seconds after the onset of the burst \cite{Fermi}. However, the constraint on the quadratic term is similar to the one derived from a gamma-ray flare of the AGN PKS 2155-304, which is located much closer, z=0.116, but was detected in gamma rays above 250\,GeV. 

Two potential problems with LIV tests with GRB and AGN are:
\begin{enumerate}
\item GRBs and AGN flares are not predictable, i.e.\ they happen at random: As a consequence it is difficult to obtain an independent confirmation of results. As a matter of fact, all LIV tests that use gamma-ray emission from astrophysical objects lack an independent confirmation.
\item The processes leading to the gamma-ray emission are not well understood. This is in particular true for GRB where it is not clear what the astrophysical origin of the gamma-ray emission is. Currently, most favored as the origin of GRB are mergers of neutron stars and the supernovae of massive stars \cite{whataregrb}. If a delay is observed it is, therefore, not clear whether the delay has to be attributed to the gamma-ray emission process or to propagation effects of the gamma rays \cite{magic}. 
\end{enumerate}

\section{LIV tests with pulsars}

Besides using gamma-ray emssion from flaring AGN and GRB, a third method to search for LIV effects is to use the pulsed gamma-ray emission from pulsars. The method was first proposed  and demonstrated with gamma-ray observations of the Crab Pulsar with EGRET by \cite{kaaret}. The idea is to time the positions of the peaks in the pulse profile and search for an energy dependent shift of the peak positions. The advantage of the method is that the precision of the peak positions depends only on the available statistics and can always be improved by accumulating more data. It is not unusual that peak positions can be determined with a precision that is better than 100 microseconds, which is a much shorter timescale than is obtained with flaring AGN, minutes, and GRB, seconds. The reasons why LIV limits with pulsars have not been competitive so far are because a) their distance: detected gamma-ray pulsars are located within a distance of a few kiloparsecs from Earth, i.e.\ are within our direct neighborhood, and because b) gamma-ray emission from pulsars was only detected up to about 10\,GeV.

The last limitation was recently overcome with the detection of the Crab pulsar above 100\,GeV by the VERITAS collaboration \cite{andrew}. The VERITAS measurement proves that the spectral cut-off is not an exponential one as it has been widely believed but that instead the pulsed emission extends to above 100\,GeV, see Figure \ref{sed}. It is not unreasonable to assume that more pulsars emit pulsed gamma rays above 100\,GeV. Being able to detect pulsed gamma rays at such high energies makes LIV tests with pulsars competitive with other methods and very attractive for several reasons:
\begin{enumerate}
\item Tests do not rely on random transient events and observational \textit{luck}. Studies can be performed in a systematic way and can be crosschecked by other experiments. 
\item Limits can be improved by observing longer.
\item Results are not based on the detection of single photons but high statistics.
\item It can be distinguished between energy delays that are intrinsic to the pulsar and gamma-ray propagation effects.
\end{enumerate}

Being able to plan a dedicated campaign and not having to wait for a random event to happen during the lifetime of an experiment is probably the most appealing and satisfying advantage from an observer's point of view. However, the advantage of being able to distinguish between source intrinsic and extrinsic effects is an important one and has not been pointed out before.

\subsection{Distinguishing between source intrinsic and extrinsic time delays}

If an energy dependent time delay is observed, the immediate question that arises is whether the delay can be attributed to LIV or if it is of astrophysical origin. If delays are observed in AGN flares or GRB observations it is very difficult, if not impossible, to make that separation. For pulsars this is different because the gamma-ray emission is emitted from within a rotating system that is slowing down with time.

If an energy dependent offset is observed in the peak positions $\Delta \Phi$ and the effect is intrinsic to the pulsar the offset will not change while the pulsar slows down. If, on the other hand, the offset is because the speed of light is energy dependent, the offset increases with time. This is because the absolute time of the propagation delay is constant $\Delta t$ but measured in the reference frame of the pulsar, i.e.\ normalized to the pulsar period, which changes with time  $P + t \cdot \dot{P}$.
\begin{equation}
\Delta \Phi(t) = \Delta t / (P + t \cdot \dot{P})
\end{equation}

\begin{figure}[ht]
  \vspace{5mm}
  \centering
  \includegraphics[width=2.6in]{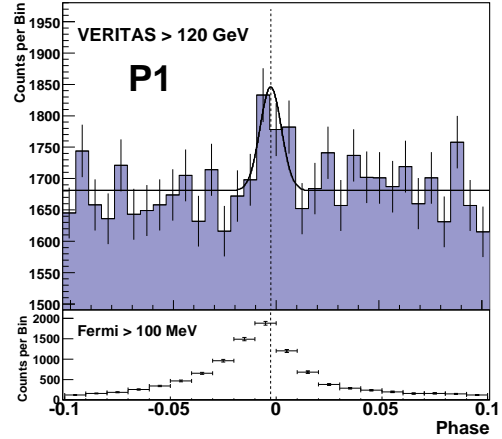}
  \caption{Exploded view of the main pulse or P1 of the Crab pulsar pulse profile. The solid line is the result of the maximum likelihood fit of the pulse profile. Figure from \cite{andrew}}
  \label{zoomP1}
 \end{figure}

For example, the Crab pulsar has a period of 30\,ms, which  gets longer at a relative rate of $\approx10^{-4}\,$ per year. The pulsar period, therefore, changes by about $3\mu$s in one year or $30\mu$s in ten years. If significant energy dependent delays smaller than these are observed it might be possible to decide whether they are due the pulsar or due to energy dependent gamma-ray propagation effects. The VERITAS collaboration determined the peak positions of the Crab pulsar above 120\,GeV with a precision of a few ten microseconds \cite{andrew}. Energy dependent delays, if they exist, are, therfore, smaller and it can be tested what their origin is.

\section{Constraining LIV with the VERITAS detection of the Crab pulsar above 120\,GeV} 

In the following we estimate how much the VERITAS detection of the Crab pulsar above 120\,GeV \cite{andrew} constrains $E_{\rm LIV}$. Figures \ref{zoomP1} and \ref{zoomP2} show parts of the pulse profile of the Crab pulsar above 120\,GeV. Fitting the pulses with an unbinned maximum likelihood method determines the position of the peaks of the pulses with an accuracy of $\delta = 2\cdot 10^{-3}$ phase units \cite{andrew}. In order to derive a limit on $E_{\rm LIV}$, the peak position have to be compared with the positions of the peaks at lower energies. A natural choice are the peak positions measured above 100\,MeV with the Fermi-LAT \cite{FermiCrab}. The peak positions agree within statistical uncertainties. In our estimate we assume that the peak positions are the same at the different energies. Furthermore, we assume that the uncertainties at 100\,MeV are much smaller than the ones at 120\,GeV. The onesided 95\% confidence level upper limit on the time difference between the peak positions between 100\,MeV and 120\,GeV is, therefore,
\begin{equation}
\Delta t_{95\%} < 1.65  \cdot \delta \cdot  P / \sqrt{  2 } < 100\,\mu\rm{s}   \,,
\end{equation} 
where P = 30\,ms is the Period of the Crab pulsar and $1/ \sqrt{  2 }$ is because the uncertainty on the position of the two peaks can be averaged.
With this limit on the time difference a limit on the energy scale of LIV can be estimated using Equation 1 and 2:

 \begin{figure}[!t]
  \vspace{5mm}
  \centering
  \includegraphics[width=2.6in]{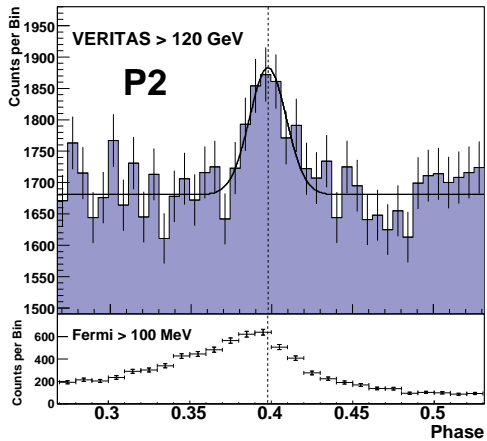}
  \caption{Exploded view of the interpulse or P2 of the Crab pulsar pulse profile. The solid line is the result of the maximum likelihood fit of the pulse profile. Figure from \cite{andrew}.}
  \label{zoomP2}
 \end{figure}

\begin{equation}
E_{\rm LIV} > \frac{2\,\rm{kpc} \cdot 120\,\rm{GeV}}{c \cdot 100\,\mu\rm{s}} \approx 3\cdot10^{17}\,\rm{GeV}
\end{equation}
for the linear term and for the quadratic term
\begin{equation}
E_{\rm LIV} > \sqrt{\frac { 3\cdot 2\,\rm{kpc}}{2\cdot c\cdot 100\,\mu\rm{s}} }\cdot 120\,\rm{GeV} \approx 7\cdot 10^9\,GeV
\end{equation}

Here we have used the conanical distance of 2\,kpc for the Crab pulsar. The limit on $E_{\rm LIV}$ in the linear term is compatible with the limit derived with the Mrk 501 flare detected by the MAGIC collaboration \cite{magic} and only one order of magnitude below the best limit derived with the observation of the PKS 2155-304 flare by H.E.S.S.\cite{HESS}. The limit on $E_{\rm LIV}$ in the  quadratic term is one order of magnitude below the best available limits \cite{HESS, Fermi}. We note that the above is only an order of magnitude estimate because it lacks the appropriate analysis of the Fermi-LAT data.
 
\section{Improvements in the future}

Although the above is only an estimate it is close to the limit that can be derived in a proper analysis. That limit can be improved by, for example, a more sophisticated analysis methods that makes use of the individual event energies.

However, an improvement by a factor of ten with the present VERITAS data is an ambitious goal and the question arises how the sensitivity can be improved further. It was already discussed above that the advantage of doing LIV tests with pulsars is the repeatability of the measurements. Deeper observations with VERITAS will certainly improve the limits but a much bigger improvement will be made with the next generation Cherenkov telescope CTA \cite{CTA} that aims at a ten times higher sensitivity than VERITAS. Observations of the Crab pulsar above 100\,GeV with CTA will improve the LIV tests in two ways:
\begin{enumerate}
\item The uncertainties on the peak positions can easily be reduced by one order of magnitude.
\item The measurements can be extended up into the TeV range, provided the spectrum extends like a power law and does not cut off earlier.
\end{enumerate}
Combining all improvements it is not unreasonable to assume that a sensitivity can be achieved, which probes LIV effects at the Planck scale by observing pulsars above 100\,GeV. While testing LIV with AGN and GRB involves a certain amount of \textit{luck} it is "guaranteed" science if done with VHE emission from pulsars.

So far the only pulsar detected above 100\,GeV is the Crab. If the photon spectra of more pulsars can be described with a power law instead of an exponential cutoff above the spectral break, it can be expected that more pulsars will be detected above 100\,GeV even with the present generation of IACT. With the prospects of detecting more pulsars above 100\,GeV additional improvements can be expected:
\begin{enumerate}
\item \textbf{Millisecond pulsars} spin about ten times faster than the Crab pulsar. The detection of a millisecond pulsar above 100\,GeV would improve limits by one order of magnitude on the linear term and give a $\sqrt{10}$ improvement on the quadratic term.
\item \textbf{Pulsars at larger distances} than the Crab pulsar if detected above 100\,GeV can improve LIV limits.
\end{enumerate}

\section{Conclusions}

The detection of the Crab pulsar above 100\,GeV shows that LIV tests with pulsars are competitive with other LIV tests. A robust estimate shows that limits on LIV can be derived with the VERITAS Crab pulsar detection that are one order of magnitude below the best available limit from an AGN observation. Pulsars also provide a unique way of separating source intrinsic effects from propagation effects, e.g.\ LIV. The limits can be improved in the future with more refined analysis methods and deeper observations. With CTA it should be possible to probe LIV with pulsar observations at the Planck mass.

\section{Acknowledgements}

I am grateful for inspiring discussions with A.\ Bouvier, A.\ Belfiore, and C.\ Otte. P.\ Kaaret was providing valuable feedback.

\clearpage

\end{document}